\documentclass[11pt, a4paper]{article}
\usepackage{ae} % or {zefonts}
\usepackage[T1]{fontenc}
\usepackage[ansinew]{inputenc}
\usepackage{amsmath, amsfonts, amssymb, diagrams, mathtools, mathabx}
\usepackage{graphicx,epsfig,skak}
\usepackage{color}
\usepackage[colorlinks]{hyperref}
\catcode`\=13
\def{$\bowtie$}
\begin{document}
\title{\textbf{Superstring BRST no-ghost theorem at arbitrary picture number}}
\author{Mykola Dedushenko\footnote{dedushenko A\_T gmail.com}\\\emph{Physics Department, Princeton University,}\\\emph{Princeton, NJ 08540, USA}}
\maketitle \abstract{A simple and self-contained treatment of the superstring BRST no-ghost theorem at non-zero momentum and arbitrary picture number is presented. We prove by applying the spectral sequence that the absolute BRST cohomology is isomorphic to two copies of the light-cone spectrum at adjacent ghost numbers. We single out a representative in each cohomology class. A non-degenerate pairing between the Fock spaces, an induced pairing on the cohomology and a Hermitian inner product on the cohomology are constructed.}
\newpage
\section{Introduction}

BRST quantization of the bosonic string theory was introduced in \cite{KatoO}, where it was shown that the string at the critical dimension (or when the total central charge vanishes) admits a BRST charge $Q_B$ satisfying $Q_B^2=0$. The way this charge acts on the Fock space of the theory (including ghost excitations) defines a BRST complex, graded by the ghost number. The space of physical states is identified as the cohomology of this complex at the ghost number $-{1\over 2}$. This approach was extensively studied in the subsequent literature. Some relevant references are: \cite{FGZ}-\cite{FurOht}.

An important question in the BRST approach is to prove that the cohomology space contains no unphysical excitations (ghost oscillators, longitudinal and timelike components of the matter oscillators) and carries a positive inner product. Either of these statements are sometimes referred to as \emph{the no-ghost theorem}. They are closely related, since unphysical excitations spoil positiveness of the inner product. We will use the name ``the no-ghost theorem'' for the first statement. Since we restrict ourselves to the case of a non-zero momentum, the statement can be formulated as an equivalence of the BRST spectrum and the light-cone spectrum. The second type of the ``no-ghost'' statement then follows easily.

The book \cite{Pol} by Polchinski presents a nice and short proof of the bosonic no-ghost theorem inspired by \cite{KatoO}. Another classical treatment is presented in \cite{FGZ} (where a vanishing theorem for the relative BRST cohomology is proven, i.e. it is shown that the relative cohomology is non-zero only at a single ghost number, and then various properties of the cohomology are elaborated). Their method actually works even when we have only $d=1$ non-compact direction  (see \cite{AsanoN}), while the approach from \cite{Pol} requires $d\ge 2$ spacetime. Note here that the zero-momentum case requires a separate discussion. In the bosonic theory it is a trivial task: there is a finite and in fact quite small number of independent on-shell states in the zero momentum Fock space. So one can study how $Q_B$ acts on these states and compute BRST cohomology just by hands, without implementing any more advanced techniques. 

The superstring BRST quantization also was extensively studied in the literature. Some references relevant to the no-ghost theorem study are: \cite{Ohta:1985af}-\cite{Kohriki:2011zza}. Some related discussion can also be found in Appendices B and C in \cite{Wit}.

There exists an approach which directly generalizes the bosonic proof from the book \cite{Pol} or from the paper \cite{KatoO} to the superstring case (see e.g. \cite{Ohta:1985af, Ito:1985qa, Itoh:1991qb, Kohriki:2010ry} ). This approach requires $d\ge 2$ as it uses the existence of two free bosons $X^+$ and $X^-$ (the light-cone directions). An approach to the superstring BRST cohomology directly generalizing the paper by Frenkel, Garland, Zuckerman \cite{FGZ} is described in \cite{LZ} (again this approach can be generalized to $d=1$, see \cite{Asano:2003jn}). However the discussion becomes more clumsy than in \cite{FGZ} because of different subtleties that appear in the superstring.

A specific feature of the superstring is the existence of the $\beta\gamma$ ghost system which is bosonic. This results in the existence of the non-equivalent representations of the $\beta\gamma$ ghosts -- the so-called different ``pictures''. Also $\beta\gamma$ zero modes (which exist only in the Ramond sector) can create certain difficulties if treated in a non-optimal way.

Below we will define a notion of the picture number $A$ (following \cite{Friedan:1985ge}). The canonical picture in the Neveu-Schwarz sector corresponds to $A=-1$, in the R sector the canonical picture is $A=-{1\over 2}$ (sometimes in the literature $A=-{3\over 2}$ is also referred to as a canonical picture, since $A=-{1\over 2}$ and $A=-{3\over 2}$ are singled out by the property that $L_0$ is bounded from below in these pictures). Most of the literature deals with the canonical pictures. There are some known results about arbitrary picture numbers. Using picture-changing operators it was proven in \cite{Horowitz:1988ip} that the absolute BRST cohomologies are isomorphic at different picture numbers. In \cite{Berkovits:1997mc} the isomorphism of relative (and semirelative) cohomologies at different picture numbers \emph{for non-zero momentum} was established (while the zero-momentum relative and semirelative cohomologies were shown to be picture-dependent, which was checked however only for canonical RR-sector pictures). Also in \cite{Kohriki:2010ry, Kohriki:2011zza} the NS-sector $A=0$ picture is discussed.

We will present here an explicit discussion of the BRST cohomology for arbitrary picture number $A$ and arbitrary sector, without using the picture-changing operators. This can be easily done with all the discussion not harder than the proof using picture-changing operators, but with an advantage of building explicit representatives of cohomology classes almost for free. However, we discuss only the non-zero momentum case. Our method is essentially the same as used in \cite{Pol} for bosonic strings but generalized to the superstring at arbitrary picture number. The full treatment of the zero-momentum superstring cohomology at arbitrary ghost number still remains an open problem.

We then construct a bilinear pairing between the $A$-picture and the $(-A-2)$-picture Fock spaces. Its connection to the inner product on the cohomology is discussed at the end.

\section{Setup}

We start with a superconformal theory with energy-momentum tensor $T_B$ and supercharge $T_F$:
\begin{align}
T_B &= T_B^d  + T_B^g\cr
T_F &= T_F^d + T_F^g,
\end{align}
where $T_B^d, T_F^d$ describe the RNS string with $10$ flat dimensions and $T_B^g, T_F^g$ are the ghost supercharge and stress tensor. The full theory has vanishing central charge and thus nilpotent BRST operator.

The $T^d+T^g$ theory is described by fields $X^{\mu}, \psi^{\mu}, b, c, \beta, \gamma$ (we can restrict ourselves without loss of generality to such case, which is an open string or a chiral part of the closed string; we will think of it as of the chiral part of the closed string). The corresponding modes satisfy (anti)commutation relations:
\begin{align}
[\alpha_m^{\mu},\alpha_n^{\nu}]&=m\delta_{m+n,0} \eta^{\mu\nu}
\cr
\{\psi^{\mu}_r,\psi^{\nu}_s\}&=\delta_{r+s,0}\eta^{\mu,\nu}
\cr
\{b_m,c_n\}&=\delta_{m+n,0}\cr
[\gamma_r,\beta_s]&=\delta_{r+s,0},
\end{align}
where $\eta=diag(-1,1,1,...,1)$ is the spacetime metric, $\mu,\nu=0\dots 9$, $m,n\in\mathbb{Z}$ and $r,s\in\mathbb{Z}+\kappa$, where we define $\kappa={1\over 2}$ in the NS sector while $\kappa=0$ in the R sector.

The corresponding space of states is defined as a Fock space. The matter field vacuum in the NS sector is defined by:
\begin{align}
\label{matter_vac}
\alpha^{\mu}_n|p\rangle&=0,\ n>0\cr
\psi^{\mu}_r|p\rangle&=0,\ r>0\cr
\alpha_0^{\mu}|p\rangle&=(2\alpha')^{1/2}p^{\mu}|p\rangle.
\end{align}
For the R-sector, we also have zero modes $\psi^{\mu}_0$ which make the Ramond matter vacuum degenerate and give it a structure of the irreducible representation of the Clifford algebra. So the R sector vacuum is spanned by the vectors $|p\rangle\otimes|u\rangle$, where $|p\rangle$ satisfies the same conditions (\ref{matter_vac}) and $|u\rangle\in \textbf{32}$ is a Dirac spinor of $Spin(9,1)$. Note that $\textbf{32}=\textbf{16}\oplus\textbf{16}'$, where $\textbf{16}$ and $\textbf{16}'$ are dual Weyl representations. Also note that all these representations -- $\textbf{32}, \textbf{16}$ and $\textbf{16}'$ -- admit real structures, i.e. Majorana conditions.

For the $bc$ ghosts we choose vacuum $|\downarrow\rangle$:
\begin{align}
b_n|\downarrow\rangle&=0,\ n\ge 0\cr
c_n|\downarrow\rangle&=0,\ n>0.
\end{align}
All other possible vacua of the $bc$-system live within the Fock space built on this single vacuum.

For the $\beta\gamma$ ghosts, the definition of vacuum is not so trivial, and actually we have infinitely many non-equivalent representations of the canonical commutation relations (see \cite{Friedan:1985ge}). Fix a number $A\in \mathbb{Z} + \kappa + {1\over 2}$. Define the $A$-picture vacuum by:
\begin{align}
\beta_r|A\rangle_{\beta\gamma}&=0,\ r\ge -A-{1\over 2}\cr
\gamma_r|A\rangle_{\beta\gamma}&=0,\ r \ge A+{3\over 2}.
\end{align}
The vacuum in the NS sector annihilated by all the positive-indexed modes corresponds to $A=-1$. The vacuum in the R sector annihilated by all the positive-indexed modes and by $\beta_0$ corresponds to the picture $A=-{1\over2}$. If we interchange the role of $\beta_0$ and $\gamma_0$ and say that the vacuum is annihilated by $\gamma_0$, we obtain an $A=-{3\over 2}$ picture. The $A=-1$ and $A=-{1\over 2}$ pictures are referred to as ``canonical pictures'' (and sometimes $A=-{3\over 2}$ as well). 

The total matter+ghosts vacuum will be denoted just by its picture number in the NS sector and also will contain a spinor state in the R sector:
\begin{align}
NS:\quad |A\rangle &= |p\rangle\otimes|\downarrow\rangle\otimes|A\rangle_{\beta\gamma}\cr
R:\quad |A,u\rangle &= |p\rangle\otimes |u\rangle\otimes|\downarrow\rangle\otimes|A\rangle_{\beta\gamma}.
\end{align}

Denote the Fock space built on this vacuum (or in the R sector on the space of vacua $|A,u\rangle$) by $\mathcal{H}^A$, where depending on the sector $A$ takes integer or half-integer values. If required, we will indicate the space-time momentum explicitly in our notation and write $\mathcal{H}^A(p)$ for the space of states. Sometimes we can write $\mathcal{H}^A_{NS}$ or $\mathcal{H}^A_R$ to emphasize the sector.

In any picture, one of the operators in a pair $(\beta_{-r},\gamma_r)$ is a creation operator and another is an annihilation operator. In particular, we have some canonical creation-annihilation choice for each pair which corresponds to the canonical pictures $A=-1$ or $A=-{1\over 2}$ (depending on the sector). In the general picture, a finite number of pairs $(\beta_{-r}, \gamma_{r})$ switch their creation-annihilation roles as compared to the $-1$ or $-{1\over 2}$ picture -- creation operators become annihilation ones and vice-versa. Let us say that $(\beta_{-r},\gamma_r)$, $r\in\mathbb{Z}+\kappa$ is a standard pair if the positive-indexed mode annihilates the vacuum and is a reversed pair if the negative-indexed-mode annihilates the vacuum. The pair $(\beta_0,\gamma_0)$ is defined to be standard if $\beta_0$ annihilates the vacuum and reversed in the opposite case. To indicate this introduce a symbol for the A-picture:
\begin{eqnarray}
\label{epsilon}
\varepsilon^A_r = \begin{cases}
r &\mbox{if } r\ne0 \mbox{ and } (\beta_{-r},\gamma_r) \mbox{ is a standard pair} \\
-r &\mbox{if } r\ne0 \mbox{ and } (\beta_{-r},\gamma_r) \mbox{ is a reversed pair}\\
1 &\mbox{if } r=0 \mbox{ and } \beta_0 \mbox{ is creation operator}\\
-1 &\mbox{if } r=0 \mbox{ and } \beta_0 \mbox{ is annihilation operator}.
\end{cases}
\end{eqnarray}

For the NS-sector $A=-1$-picture, where all positive-indexed modes annihilate the vacuum, we have $\varepsilon^{-1}_r=r$ for all $r$. For the R-sector $-{1\over 2}$-picture we have $\varepsilon^{-1/2}_r=r$ for $r\ne 0$ and $\varepsilon^{-1/2}_0=-1$. In going to different picture, every time when operators in the pair $(\beta_{-r},\gamma_r)$ switch their creation-annihilation roles, the number $\varepsilon_r$ changes its sign. For later reference denote by $K^A$ the set of such $r$'s, for which $\varepsilon^A_r$ changes the sign as compared to the $-1$-picture in the NS and to the $-{1\over 2}$-picture in the R-sector:
\begin{align}
NS:\ K^A&=\{r\in\mathbb{Z}+{1\over 2}: \varepsilon^A_r=-\varepsilon^{-1}_r\}\cr
R:\ K^A&=\{r\in\mathbb{Z}: \varepsilon^A_r = -\varepsilon^{-1/2}_r\}.
\end{align}

We can explicitly describe these sets:
\begin{align}
NS:\quad\quad K^A&=\begin{cases}\left\{{1\over 2},{3\over 2},...,A+{1\over 2}\right\}, &\mbox{if } A\ge 0\\
\left\{A+{3\over 2}, A+{5\over 2}, ..., -{1\over 2}\right\}, &\mbox{if } A\le -2\\
\varnothing, &\mbox{if } A=-1
\end{cases}\cr
R:\quad\quad K^A &= \begin{cases}\left\{1,2,...,A+{1\over 2}\right\}, &\mbox{if } A\ge-{1\over 2}\\
\left\{A+{3\over 2},A+{5\over 2},...,0\right\}, &\mbox{if } A\le-{3\over 2}\\
\varnothing, &\mbox{if } A=-{1\over 2}.\end{cases}
\end{align}

From the conformal properties one can derive for the $\beta\gamma$ ghost number current in the A-picture:
\begin{equation}
j = -{}_{\bullet}^{\bullet}\beta\gamma{}_{\bullet}^{\bullet} = -\sum_{r,s\in\mathbb{Z}+\nu}{:\beta_r\gamma_s:\over z^{r+s+1}}+{A\over z},
\end{equation}
where ${}_{\bullet}^{\bullet}\dots{}_{\bullet}^{\bullet}$ is conformal normal-ordering (subtracting of a simple pole, which is equivalent to the normal-ordering with respect to the $A=0$ vacuum), while $:\dots:$ is normal-ordering with respect to the $A$-picture vacuum (we do not write $A$ explicitly). We will use $:..:$ normal-ordering later in this paper.  The ghost number is conventionally defined using the cylindrical coordinates $w$, which are connected with coordinates $z$ on the plane by the conformal transformation $z=e^{-iw}$. Because of the non-tensorial behavior of the ghost current, this gives an additional shift $+1$ to the ghost number. So the ghost number of the vacuum gets contribution $A+1$ from $\beta\gamma$ and $-{1\over 2}$ from $bc$ ghosts, the total being $A+{1\over 2}$.

Introduce a light-cone basis:
\begin{align}
\alpha^{\pm}_n &= {\alpha^0_n \pm \alpha^1_n\over\sqrt{2}}\cr
\psi^{\pm}_r &= {\psi^0_r \pm \psi^1_r\over\sqrt{2}},
\end{align}
for which we have:
\begin{align}
[\alpha^+_m,\alpha^-_n]&=-m\delta_{m+n,0}\cr
\{\psi^+_r,\psi^-_s\}&=-\delta_{r+s,0}.
\end{align}
Later we will refer to excitations of the $\pm$ modes as the light-cone excitations.

In the BRST quantization, we define a space of physical states as the cohomology of the BRST operator $Q_B$
\begin{eqnarray}
Q_B = \sum_m c_{-m}L^{matter}_m + \sum_{r}\gamma_{-r}G_r^{matter} - \sum_{m,n}{1\over 2}(n-m):b_{-m-n}c_mc_n:+\cr
+\sum_{m,r}\left[ {1\over 2}(2r-m):\beta_{-m-r}c_m\gamma_r:-:b_{-m}\gamma_{m-r}\gamma_r: \right] + a^g c_0.
\end{eqnarray}

The matter superconformal generators $L^{matter}_m,G^{matter}_r$ in either sectors are given by:
\begin{align}
L^{matter}_m &= {1\over 2}\sum_{n\in \mathbb{Z}}:\alpha^{\mu}_{m-n}\alpha_{\mu n}:+{1\over 4}\sum_{r\in \mathbb{Z}+\nu}(2r-m):\psi^{\mu}_{m-r}\psi_{\mu r}: + a^m\delta_{m,0}\cr
G^{matter}_r &= \sum_{n\in \mathbb{Z}}\alpha^{\mu}_n\psi_{\mu r-n}.
\end{align}

We have:
\begin{align}
\{Q_B,b_n\}&=L_n\cr
[Q_B,\beta_r]&=G_r,
\end{align}
where $L_n$ and $G_r$ are \emph{matter + ghost} superconformal generators. Because of $L_0=\{Q_B,b_0\}$ and $[Q_B,L_0]=0$ one can look for cohomologies of $Q_B$ in the eigenspaces of $L_0$. Moreover, from here it follows that all the cohomologies are within the $L_0=0$ subspace, but we'll re-derive it later.
 
\section{Light-cone grading and filtration}

Define a light-cone grading by the charge:
\begin{equation}{
N^{lc} = \sum_{m\in \mathbb{Z}\backslash 0} {1\over m}:\alpha^+_{-m}\alpha^-_m:+\sum_{r\in\mathbb{Z}+\nu}:\psi^+_{-r}\psi^-_r:.
}
\end{equation}

This charge measures the number of excitations of the ``--'' modes minus the number of excitations of the ``+'' modes. Note that in the NS sector there are no zero-modes, while in the R-sector a zero mode excitation $\psi^+_0$ appears among the ``+''-excitations. Assume that we are working in the Fock space with a non-zero momentum $k$. Choose a Lorentz frame in which $k^+\ne 0$, so $\alpha_0^+=(2\alpha')^{1/2}k^+\ne 0$.

\textbf{Lemma 1:} $Q_B$ decomposes under this grading into:
\begin{equation}{
Q_B = Q_{-1} + Q_0 + Q_{+1},
}
\end{equation}
where 
\begin{equation}{
[N^{lc},Q_j]=jQ_j,
}
\end{equation}
and so $Q_1^2 = Q_{-1}^2 = \{Q_1,Q_0\}=\{Q_{-1},Q_0\}=0$. Each $Q_j$ has ghost number $+1$ and we have:
\begin{equation}{
\label{well_def}
\{Q_1,b_0\}=[Q_1,L_0]=0.
}
\end{equation}
An expression for $Q_1$ in both sectors is
\begin{equation}{
Q_1 = -\alpha_0^+\left( \sum_{m\in \mathbb{Z}\backslash 0}c_{-m}\alpha_{m}^- + \sum_{r\in\mathbb{Z}+\nu}\gamma_{-r}\psi^-_{r} \right).
}
\end{equation}
\textbf{Proof:} All the statements can be checked by direct calculation. For (\ref{well_def}) we can also give a short derivation: one can check that $[N^{lc},L_0]=[N^{lc},b_0]=0$; so taking the $N^{lc}=1$ component of identities $[Q_B,L_0]=0$ and $\{Q_B,b_0\}=L_0$ we get $[Q_1,L_0]=0$ and $\{Q_1,b_0\}=0$. $\square$ 

We also introduce a filtration degree:
\begin{equation}{
\label{filtr_deg}
N^f = N^g - N^{lc},
}
\end{equation}
where $N^g$ is the ghost number. Note that $[N^f,Q_j]=(1-j)Q_j$, and so the BRST operator preserves the following filtration on the full Fock space $\mathcal{H}^A_{NS, R}$:
\begin{equation}{
\mathcal{H}^{(p)}=\{\psi\in\mathcal{H}_{NS,R}^A: N^f(\psi)\ge p\}.
}
\end{equation}
This suggests that we can use spectral sequences here. Those who are not familiar with spectral sequences can find a concise introduction in \cite{Lang}. Using physical terminology, the main idea can be summarized as follows: apply perturbation theory to compute the cohomology of $Q_B=Q_1 + Q_0 + Q_{-1}$. First, we find the cohomology of the operator $Q_1$ and then look for the higher corrections. The spectral sequence is the most effective when this perturbative procedure converges after the finite number of steps (in our case it converges after the first step). 

The graded components of the associated graded space (which forms the zeroth term of the spectral sequence) are identified as subspaces of the definite filtration degree:
\begin{equation}{
E_0^p = \mathcal{H}^{(p)}/\mathcal{H}^{(p+1)}\cong\{\psi\in\mathcal{H}^A_{NS,R}: N^f(\psi)=p\}.
}
\end{equation}
We do not indicate a sector and picture number for $E_0$ and $\mathcal{H}^{(p)}$ to make our notation less clumsy. Note:
\begin{equation}{
\bigoplus_{p}E_0^p \cong E_0 \cong \mathcal{H}_{NS,R}^A },
\end{equation}
and $Q_B$ induces a differential on $E_0$, which is precisely the $Q_1$ operator described above. So now we have to compute the cohomology of $Q_1$.

\section{Cohomology of $Q_1$}

Introduce an operator:
\begin{equation}{
\label{R_operator}
R={1\over\alpha_0^+}\left( \sum_{m\ne 0}b_{-m}\alpha^+_{m} + \sum_{r\in \mathbb{Z}+\nu}\varepsilon_r^A\beta_{-r}\psi_r^+ \right),
}
\end{equation}
where we used the symbol $\varepsilon_r^A$ defined in (\ref{epsilon}). One can check that this satisfies $[R,L_0]=\{R,b_0\}=0$. For the ghost number and for the light-cone charge of this operator we have:
\begin{align}
\label{R_gh}
[N^{gh},R]&=-R\\
\label{R_lc}
[N^{lc},R]&=-R,
\end{align}
which will prove to be useful later.

Define the operator:
\begin{eqnarray}
S=\{Q_1,R\}=\sum_{n=1}^{\infty}(nc_{-n}b_n + nb_{-n}c_n - \alpha^+_{-n}\alpha^-_n - \alpha^-_{-n}\alpha^+_n)\cr +\sum_{r\in\mathbb{Z}+\nu}(\varepsilon_r^A\beta_{-r}\gamma_r - \varepsilon_r^A \psi^-_{-r}\psi^+_{r}).
\end{eqnarray}
Define also a quantum number $N_{\beta\gamma r}$ for $r>0$ -- it counts the number of excitations of the $(\beta_{-s},\gamma_s)$ oscillators with $|s|=r$. That is, if $\beta_r,\gamma_r$ are annihilation operators, then $N_{\beta\gamma r}=\beta_{-r}\gamma_r - \gamma_{-r}\beta_r$; if $\beta_r,\beta_{-r}$ are annihilation operators, then $N_{\beta\gamma r}=-\gamma_{-r}\beta_r - \gamma_{r}\beta_{-r}$; and if $\gamma_r, \gamma_{-r}$ are annihilation operators, then $N_{\beta\gamma r}=\beta_{-r}\gamma_r + \beta_{r}\gamma_{-r}$ ($\beta_{-r}$ and $\gamma_{-r}$ cannot be annihilation operators at once for the vacuum $|A\rangle$). For $r=0$ we have: if $\beta_0$ is an annihilation operator, then $N_{\beta\gamma 0}=-\gamma_0\beta_0$, if $\beta_0$ is a creation operator, then $N_{\beta\gamma 0}=\beta_0\gamma_0$.

We also wish to reverse the creation-annihilation roles for the operators $\psi_r^+, \psi_{-r}^-$ every time that the pair $(\beta_{-r},\gamma_r)$ is reversed, i.e. when $r\in K^A$. Due to the fermionic nature of $\psi$'s, this actually means that we consider all the states as excitations above a new vacuum\footnote{In the NS sector at $A=-1$ we do not change the vacuum, and we have $R={1\over\alpha_0^+}\left( \sum_{m\ne 0}b_{-m}\alpha^+_{m} + \sum_{r\in \mathbb{Z}+\nu}r\beta_{-r}\psi_r^+ \right)$. When we go to a different picture, some of the pairs $(\beta_{-r},\gamma_r)$ reverse their creation-annihilation roles, and for such $r$'s we should flip the sign in front of $\beta_{-r}$ in the definition of the operator $R$. This operation is encoded in the definition of $\varepsilon_r^A$. Also for such $r$'s we switch the creation-annihilation roles of $\psi_r^+$ and $\psi_{-r}^-$ -- this is encoded in our definition of the new vacuum $|\widetilde{A}\rangle$. In the R-sector everything is the same except that at $A=-{1\over 2}$ we take $R={1\over\alpha_0^+}\left( \sum_{m\ne 0}b_{-m}\alpha^+_{m} + \sum_{r\in \mathbb{Z}+\nu}r\beta_{-r}\psi_r^+  - \beta_0\psi_0^+\right).$}:
\begin{align}
\label{truevac}
NS:\quad |\widetilde{A}\rangle&=\prod_{\begin{matrix}
r> 0\cr
r\in K^A
\end{matrix}}\psi_{-r}^- \prod_{\begin{matrix}
r< 0\cr
r\in K^A
\end{matrix}}\psi_{r}^+ |A\rangle\\
\label{truevac_R}
R:\quad |\widetilde{A,u}\rangle&=\prod_{\begin{matrix}
r> 0\cr
r\in K^A
\end{matrix}}\psi_{-r}^- \prod_{\begin{matrix}
r\le 0\cr
r\in K^A
\end{matrix}}\psi_{r}^+ |A,u\rangle, \ \mbox{where } \psi_0^- u=0,
\end{align}
where for definiteness we fix the ordering of $\psi$'s in such a way that the absolute values of mode numbers grow from the right to the left. Note also that with this definition, in the R sector for $A\ge -{1\over 2}$, we have $\psi_0^-|\widetilde{A,u}\rangle=0$, while for $A\le -{3\over 2}$, we have $\psi_0^+|\widetilde{A,u}\rangle=0$. This reduces the dimension of the space of spinors by the fator of 2.

Analogously define the operator $N^{\pm}_{\psi r}$ that measures the number of $\psi^{\pm}_s$ excitations with $|s|=r$ \textbf{above the vacuum $|\widetilde{A}\rangle$ or $|\widetilde{A,u}\rangle$}.

The symbol $\varepsilon^A_r$ is defined in precisely such a way that the expression for $S$ can be rewritten in each sector in a simple way using the numbers $N_{\beta\gamma r}$ and $N_{\psi r}^{\pm}$:
\begin{align}
\label{S_operator}
NS:\quad S&=\sum_{n=1}^{\infty}n(N_{cn}+N_{bn}+N^+_{Xn} + N^-_{Xn})+\sum_{r\in\mathbb{Z}+{1\over 2}, r>0}r(N_{\beta\gamma r}+N^{\pm}_{\psi r})\cr
R:\quad S&=\sum_{n=1}^{\infty}n(N_{cn}+N_{bn}+N^+_{Xn} + N^-_{Xn})+\sum_{r=1}^{\infty}r(N_{\beta\gamma r}+N^{\pm}_{\psi r})+N_{\beta\gamma 0}+N^{\pm}_{\psi 0}.
\end{align}

This operator commutes with $Q_1$ due to the Jacobi identity. It also commutes with $b_0$ and $L_0$.

\textbf{Definition:} let $\mathcal{H}^{\perp}_{A,\downarrow}$ be a Fock space built over the state $|\widetilde{A}\rangle$ in the NS sector or over the space of vacuums $|\widetilde{A,u}\rangle, u\in\textbf{32}, \psi_0^- u=0$ in the R sector by exciting $\psi_{-r}^i$ and $\alpha_{-n}^i$ modes with $i=2\dots9$. Let $\mathcal{H}^{\perp}_{A,\uparrow}$ be a Fock space built over the state $c_0|\widetilde{A}\rangle$ in the NS sector or over the space of vacuums $c_0|\widetilde{A,u}\rangle, u\in\textbf{32}, \psi_0^- u=0$ in the R sector by exciting the same modes.

Recall that $Q_1$ defines a complex $(\mathcal{H}^A,Q_1)$ (we do not indicate sector here), and we need its cohomology:

\textbf{Lemma 2:} $H^*\big((\mathcal{H}^A,Q_1)\big)\cong\ker S= \mathcal{H}^{\perp}_{A,\downarrow}\oplus\mathcal{H}^{\perp}_{A,\uparrow}$. 

\textbf{Proof:} Since $S$ commutes with $Q_1$, it is enough to compute the $Q_1$ cohomology in each eigenspace of $S$. Since $S=\{Q_1,R\}$, we have $H^*\big((\mathcal{H}^A,Q_1)\big)\subset \ker S$.

Now consider the kernel of $S$. From (\ref{S_operator}), it follows that the kernel of S has no light-cone (i.e. $\pm$) excitations above the vacuum $|\widetilde{A}\rangle$ or $|\widetilde{A,u}\rangle$. Also it has no ghost excitations except possibly the $c_0$ excitation, which does not change the $S$ eigenvalue. This means that actually:
$$
\ker S= \mathcal{H}^{\perp}_{A,\downarrow}\oplus\mathcal{H}^{\perp}_{A,\uparrow},
$$
where the spaces on the r.h.s. were defined above. The first space has no ghost excitations, and so its ghost number is $A+{1\over 2}$. The second space has only $c_0$ excitation, and so its ghost number is $A+{3\over 2}$:
\begin{align}
N^{gh}\mathcal{H}^{\perp}_{A,\downarrow}&=\left(A+{1\over 2}\right)\mathcal{H}^{\perp}_{A,\downarrow}\cr
N^{gh}\mathcal{H}^{\perp}_{A,\uparrow}&=\left(A+{3\over 2}\right)\mathcal{H}^{\perp}_{A,\uparrow}.
\end{align}
Also these spaces are connected by:
\begin{equation}
\mathcal{H}^{\perp}_{A,\downarrow}=b_0\mathcal{H}^{\perp}_{A,\uparrow}.
\end{equation}

Notice that for $|\psi\rangle\in \mathcal{H}^{\perp}_{A,\uparrow}$, as a consequence of $[Q_1,S]=0$:
$$
S Q_1|\psi\rangle=Q_1 S|\psi\rangle = 0,
$$
so $Q_1|\psi\rangle\in\ker S$. But such a state has ghost number $A+{5\over 2}$, which is consistent only if $Q_1|\psi\rangle=0$. So $Q_1$ annihilates the full subspace $\mathcal{H}^{\perp}_{A,\uparrow}$. For the subspace $\mathcal{H}^{\perp}_{A,\downarrow}$ we have:
\begin{equation}
\label{onshell}
Q_1\mathcal{H}^{\perp}_{A,\downarrow}=Q_1 b_0 \mathcal{H}^{\perp}_{A,\uparrow} = -b_0 Q_1\mathcal{H}^{\perp}_{A,\uparrow}=0.
\end{equation}
This shows that $Q_1$ annihilates both $\mathcal{H}^{\perp}_{A,\downarrow}$ and $\mathcal{H}^{\perp}_{A,\uparrow}$, and so all the states in $\ker S$ are $Q_1$-closed. Thus there are no $Q_1$-exact states there, which proves the Lemma.$\square$

\textbf{Remark:} The kernel of $S$ has a definite value of $N^{lc}$, which depends on A. One can easily find by looking at the vacuum $|\widetilde{A}\rangle$ for the NS sector that it has $N^{lc}=A+1$, while for R-sector $|\widetilde{A,u}\rangle$ has $N^{lc}=A+{1\over 2}$.

\section{Cohomology of $Q_B$}
Above, we have computed the cohomology of $Q_1$, i.e. the cohomology of the zeroth term of the spectral sequence. By definition this gives the first term $E_1$. Lemma 2 shows that it is non-zero only at ghost numbers $A+{1\over 2}$ and $A+{3\over 2}$. In terms of the filtration degree, it is non-zero only at degrees $\alpha\pm{1\over 2}$, where $\alpha=0$ in the NS sector and $\alpha={1\over 2}$ in the R sector. So the only non-trivial components of the first term of the spectral sequence are (cohomology is graded by the filtration degree here):
\begin{align}
E_1^{\alpha\pm{1\over 2}}&=H^{\alpha\pm{1\over 2}}\left((\mathcal{H}^A,Q_1)\right)\cr
E_1^{\alpha-{1\over 2}}&\cong \mathcal{H}^{\perp}_{A,\downarrow}\cr
E_1^{\alpha+{1\over 2}}&\cong \mathcal{H}^{\perp}_{A,\uparrow}.
\end{align} 

As follows directly from the construction, the differential on $E_1$ is the one induced by $Q_0$ on the cohomology of $Q_1$ (still denote this operator by $Q_0$). So already at the first term we are left just with the following:
\begin{equation}{
\label{pair}
\begin{diagram}
0        &\rTo &E_1^{\alpha-1/2}&\rTo^{Q_0} & E_1^{\alpha+1/2}&\rTo & 0.\\
\end{diagram}
}
\end{equation}

Next we should notice that the operator $Q_0$ can be rewritten in a useful way:
\begin{equation}
Q_0=\widetilde{Q}_0 + c_0 L_0.
\end{equation}
By Lemma 1, $\{Q_1,Q_0\}=0$, and since $\{Q_1,c_0\}=[Q_1,L_0]=0$, this implies $\{Q_1, \widetilde{Q}_0\}=0$. Thus, $\widetilde{Q}_0$ acts on the cohomology of $Q_1$. Next, notice that $\widetilde{Q}_0$ commutes with $b_0$ (and so does not excite the $c_0$ mode). But since $\widetilde{Q}_0$ still increases the ghost number by one, it excites some other (non-zero) ghost mode. Thus, the range of the operator $\widetilde{Q}_0$ lies within the subspace where $S$ is invertible. But we know that there are no cohomologies in such subspace. In other words, $\widetilde{Q}_0$ maps into the subspace which contains no cohomologies. This shows that $\widetilde{Q}_0$ is trivial on cohomology. So (\ref{pair}) actually takes the form:
\begin{equation}{
\begin{diagram}
0        &\rTo &E_1^{\alpha-1/2}&\rTo^{c_0 L_0} & E_1^{\alpha+1/2}&\rTo & 0.\\
\end{diagram}
}
\end{equation}
So now we see that off-shell this complex describes just an isomorphism. Thus its cohomology vanishes off-shell, and the spectral sequence abuts to zero -- there are no BRST cohomology off-shell.

On-shell, $L_0=0$, and so the spectral sequence stabilizes: the only non-zero terms are $E_{\infty}^{\alpha\pm 1/2}\cong E_1^{\alpha\pm 1/2}\cap \ker L_0$. Since the spectral sequence abuts to cohomology of the total BRST operator $Q_B$, this implies that the BRST cohomology is isomorphic to the on-shell part of the $Q_1$ cohomology:
\begin{align}
\label{BRST_cohom}
H^*\left((\mathcal{H}^A,Q_B)\right)&\cong H^*\left((\mathcal{H}^A,Q_1)\right)\cap \ker L_0\cr
&\cong \ker S\cap \ker L_0 = (\mathcal{H}^{\perp}_{A,\downarrow}\oplus \mathcal{H}^{\perp}_{A,\uparrow})\cap\ker L_0.
\end{align}

So if we grade the cohomology by the ghost number $N^g$ we see that for the picture number $A$, we have two copies of the light-cone spectra at ghost numbers $(A+1)\pm{1\over 2}$.

\section{Explicit realization}

Although we have just proven by means of the spectral sequence that $Q_B$ cohomology is isomorphic to $(\mathcal{H}^{\perp}_{A,\downarrow}\oplus \mathcal{H}^{\perp}_{A,\uparrow})\cap\ker L_0$, this isomorphism doesn't respect the action of $Q_B$. Sometimes it is useful to pick a representative from each cohomology class and form another vector space isomorphic to the cohomology, each vector of which represents some BRST class in the bigger Fock space. We will now proceed to do so using the method described in \cite{Pol}, where the bosonic string case was considered.
 
Define an operator $U$ by:
\begin{equation}{
S+U = \{Q_B,R\},
}
\end{equation}
where now the operator $U=U_{-1}+U_{-2}$ with $U_{-1}=\{Q_0,R\}$ and $U_{-2}=\{Q_{-1},R\}$ has the following properties (because of (\ref{R_gh}),(\ref{R_lc})):
\begin{align}
[N^{gh},U]&=0\\
[N^{lc},U_{-1}]&=-U_{-1}\\
[N^{lc},U_{-2}]&=-2U_{-2}.
\end{align}
These imply that in terms of the $N^{lc}$ eigenvalues, $S$ is a diagonal operator, while $U$ is given by a lower triangular matrix. So one can prove the Lemma:

\textbf{Lemma 3:} $\ker (S+U) \cong \ker S$. Moreover $\ker (S+U)= \tilde{\mathcal{H}}^{\perp}_{A+1/2}\oplus \tilde{\mathcal{H}}^{\perp}_{A+3/2}$, where $\tilde{\mathcal{H}}^{\perp}_{A+1/2}=\Omega \mathcal{H}^{\perp}_{A\downarrow}$, $\tilde{\mathcal{H}}^{\perp}_{A+3/2}=\Omega \mathcal{H}^{\perp}_{A\uparrow}$ and
$$
\Omega = 1 - S^{-1}U + (S^{-1}U)^2 - \dots 
$$

\textbf{Proof:}
The operators $S$ and $U$ commute with $N^{gh}$ and with $L_0$ (since $Q_B$ and $R$ commute with $L_0$), and $N^{gh}$ and $L_0$ commute within each other. Thus it is enough to prove the Lemma in a subspace with fixed values of $L_0$ and $N^{gh}$. Such a subspace is finite dimensional.

So we consider $S$ as a finite diagonal matrix and $U$ as a finite lower-triangular matrix. Then the kernel of $S+U$ is not bigger then the kernel of $S$. In our case these kernels are actually equal in size and we can build an explicit isomorphism $\Omega$. For any $|\psi_0\rangle\in\ker S$ we take:
\begin{equation}{
\label{isom}
|\psi\rangle = \Omega|\psi_0\rangle = |\psi_0\rangle-S^{-1}U|\psi_0\rangle + S^{-1}US^{-1}U|\psi_0\rangle-\dots
}
\end{equation}
Note that this expression is well-defined: first of all, $U$ lowers $N^{lc}$, and so $S^{-1}$ always acts on the states where $S$ is actually invertible. Second, since $S^{-1}U$ is lower triangular (in the finite subspace), it is nilpotent, and so the sum above actually contains only finite number of terms.

This map obviously establishes the required isomorphism of kernels: $\ker(S+U)=\Omega\ker S$. Since we know that
$$
\ker S= \mathcal{H}^{\perp}_{A,\downarrow}\oplus\mathcal{H}^{\perp}_{A,\uparrow},
$$
this immediately finishes the proof. $\square$

\textbf{Remark:} Notice that $[b_0,S]=0$, and $[b_0,U]=[b_0,S+U]=[b_0,\{Q_B,R\}]=[L_0,R]=0$. Thus $[b_0,\Omega]=0$ and we have:
\begin{equation}{
\tilde{\mathcal{H}}^{\perp}_{A+{1\over 2}}=\Omega\mathcal{H}^{\perp}_{A\downarrow}= \Omega b_0\mathcal{H}^{\perp}_{A\uparrow}= b_0\Omega \mathcal{H}^{\perp}_{A\uparrow}=b_0 \tilde{\mathcal{H}}^{\perp}_{A+{3\over 2}}.
}
\end{equation}

However $[c_0,U]\ne 0$ and so $\tilde{\mathcal{H}}^{\perp}_{A+{3\over 2}}\ne c_0 \tilde{\mathcal{H}}^{\perp}_{A+{1\over 2}}$.

Now we notice that by definition and by the Jacobi identity $[Q_B,S+U]=0$. So the cohomology of $Q_B$ is represented by the subspace of $\ker(S+U)$ because of $S+U=\{Q_B,R\}$.

From the Lemma 3, we know that $\ker(S+U)=\tilde{\mathcal{H}}^{\perp}_{A+1/2}\oplus \tilde{\mathcal{H}}^{\perp}_{A+3/2}$ and from the Remark we know that $\tilde{\mathcal{H}}^{\perp}_{A+1/2}=b_0 \tilde{\mathcal{H}}^{\perp}_{A+3/2}$. Also, since $\Omega$ preserves ghost number, we know that $N^{gh}\tilde{\mathcal{H}}^{\perp}_{A+1/2}=(A+{1\over 2})\tilde{\mathcal{H}}^{\perp}_{A+1/2}$ and $N^{gh}\tilde{\mathcal{H}}^{\perp}_{A+3/2}=(A+{3\over 2})\tilde{\mathcal{H}}^{\perp}_{A+3/2}$. This allows us to act as in Lemma 2. Since $Q_B$ raises ghost number, we deduce that:
$$
Q_B \tilde{\mathcal{H}}^{\perp}_{A+3/2} = 0
$$
because $Q_B$ preserves $\ker (S+U)$, which has ghost number not higher than $A+3/2$. For $\tilde{\mathcal{H}}^{\perp}_{A+1/2}$ we have:
\begin{equation}{
\label{offshell}
Q_B\tilde{\mathcal{H}}^{\perp}_{A+1/2}=Q_B b_0 \tilde{\mathcal{H}}^{\perp}_{A+3/2} = \{Q_B,b_0\}\tilde{\mathcal{H}}^{\perp}_{A+3/2}=L_0 \tilde{\mathcal{H}}^{\perp}_{A+3/2}.
}
\end{equation}
This shows that off-shell $Q_B$ doesn't annihilate $\tilde{\mathcal{H}}^{\perp}_{A+1/2}$, but instead maps it isomorphically onto $\tilde{\mathcal{H}}^{\perp}_{A+3/2}$ (its off-shell part), thus making all the states there BRST-exact. So we obtain one more time that there is no BRST-cohomology off-shell.

However on-shell (\ref{offshell}) shows that $Q_B$ annihilates $\tilde{\mathcal{H}}^{\perp}_{A+1/2}\cap\ker L_0$. This means that:
$$
H^*\big((\mathcal{H}^A,Q_B)\big)\cong\ker (S+U)\cap\ker L_0= (\tilde{\mathcal{H}}^{\perp}_{A+1/2}\oplus \tilde{\mathcal{H}}^{\perp}_{A+3/2})\cap\ker L_0.
$$

\textbf{Remark:} Since $b_0\tilde{\mathcal{H}}^{\perp}_{A+1/2}=0$, the $Q_B$ cohomologies consist of just two copies of the relative cohomology at ghost numbers $A+{1\over 2}$ and $A+{3\over 2}$ (relative cohomology is the cohomology of the subcomplex annihilated by $b_0$). These are again the two copies of the light-cone spectra discussed in the previous section.

\section{Non-degenerate bilinear pairing of Fock spaces}
A two-point function on a sphere induces a canonical non-degenerate bilinear pairing between $A$-picture and $(-A-2)$-picture Hilbert spaces of opposite space-time momenta:
\begin{equation}
\label{pairing_H}
\mathcal{H}^A(p)\times \mathcal{H}^{-A-2}(-p) \to \mathbb{C},
\end{equation}
which is a tree amplitude with two incoming string states. The anomaly in the ghost number current of the $\beta\gamma$-system requires that the total picture number of the vertex operators be $-2$, implying that the $A$-picture space of states can pair non-trivially only to the $(-A-2)$-picture space. Since both strings are \emph{incoming}, they must have opposite space-time momenta. One can also show separately (like in \cite{LZ}) that there is no consistent way to introduce bilinear pairing between any other spaces.

Let the sphere be covered by two maps with holomorphic coordinates $z$ and $w$, respectively, with gluing function $w={1\over z}$. Consider states $\psi \in \mathcal{H}^A$, $\phi \in \mathcal{H}^{-A-2}$. We then take the corresponding vertex operators $\Psi_w(w)$, $\Phi_z(z)$, the first written in terms of the $w$ coordinate, and the second one in terms of the $z$ coordinate. Then we want the following property to hold for the pairing (\ref{pairing_H}) denoted by $(\ ,\ )$:
\begin{equation}{
(\psi,\phi)\ \propto\ \langle\Psi_w(w=0) \Phi_z(z=0)\rangle,
}
\end{equation}
where $\langle\dots\rangle$ denotes the two-point function on the sphere\footnote{There exists another definition in the literature (e.g. \cite{DiF}), where both fields are taken in the same coordinate patch, and for the field $\Psi(z)$ of conformal dimension $h$ the pairing is $\lim_{z\to\infty} z^{2h}\langle \Psi(z)\Phi(0)\rangle$. This definition is equivalent to ours if the field $\Psi(z)$ is primary. However, if the field is not primary, the definitions are not equivalent, and our definition makes more sense for our purposes.}. We will fix the normalization soon.

This pairing is obviously Lorentz-invariant. For completeness of the discussion, we will describe it explicitly in our fixed Lorentz frame.

Note that the space $\mathcal{H}^A$ has a natural $\mathbb{Z}_2$ grading by the fermion number (which is different from the GSO grading). One can think of this grading as being induced by the state-operator correspondence from the $\mathbb{Z}_2$ grading on the space of fields, where it describes the world-sheet statistics of the vertex operators.  Let us denote the $\mathbb{Z}_2$-grading of the general state $|\phi\rangle$ of definite degree by $(-1)^{\widetilde{\phi}}$. See Appendix A for details and subtleties about the $\mathbb{Z}_2$-grading on $\mathcal{H}^A$.

\subsection{Conjugating with respect to the pairing}

Suppose we have a weight $h$ holomorphic primary field $O(z)=\sum_n {O_n\over z^{n+h}}$ of definite $\mathbb{Z}_2$-degree. We may want to consider an expression $(|\psi\rangle,O_n |\phi'\rangle)$ and figure out how the conjugation with respect to our pairing works. To do this, we represent $O_n =\oint {dz\over 2\pi i} z^{n+h-1} O(z)$, use the fact that that the two regions of the sphere are glued by $w={1\over z}$ and flip the contour over the sphere. As a result:
\begin{equation}{
(|\phi\rangle,O_n |\phi'\rangle)=(-1)^h (-1)^{\widetilde{O}\widetilde{\phi}}(O_{-n}|\phi\rangle,|\phi'\rangle),
}
\end{equation}
where $\widetilde{O}$ is $0$ or $1$ depending on whether the field $O(z)$ is bosonic or fermionic. Using this, we can easily find:
\begin{align}
\label{conj_prop}
(|\phi\rangle,\alpha^{\mu}_n|\phi'\rangle)&=-(\alpha^{\mu}_{-n}|\phi\rangle,|\phi'\rangle), &(|\phi\rangle,\beta_r|\phi'\rangle)&=-i(\beta_{-r}|\phi\rangle,|\phi'\rangle)\cr
(|\phi\rangle,\gamma_r|\phi'\rangle)&=-i(\gamma_{-r}|\phi\rangle,|\phi'\rangle), &(|\phi\rangle,\psi^{\mu}_r|\phi'\rangle)&=i(-1)^{\widetilde{\phi}}(\psi^{\mu}_{-r}|\phi\rangle,|\phi'\rangle)\cr
(|\phi\rangle,b_n|\phi'\rangle)&=(-1)^{\widetilde{\phi}}(b_{-n}|\phi\rangle,|\phi'\rangle), &(|\phi\rangle,c_n|\phi'\rangle)&=-(-1)^{\widetilde{\phi}}(c_{-n}|\phi\rangle,|\phi'\rangle).
\end{align}
For $Q_B$, which is an integral of the BRST current, one then has:
\begin{equation}{
\label{q_dual}
(|\phi\rangle,Q_B|\phi'\rangle)+(-1)^{\widetilde{\phi}}(Q_B|\phi\rangle,|\phi'\rangle)=0.
}
\end{equation}

\subsection{Fixing normalization}
In the NS sector, it is a trivial task to fully fix our pairing. We want $(|A\rangle,c_0|-A-2\rangle)=\pm 1$, where $c_0$ insertion is required to saturate the $c$-field zero modes. We fix the sign by the requirement:
\begin{equation}
\label{nsfix}
NS:\quad (|\widetilde{A}\rangle,c_0|\widetilde{-A-2}\rangle)=1.
\end{equation}

In the Ramond sector the condition is 
\begin{equation}
\label{rfix}
R:\quad (|\widetilde{A,u}\rangle, c_0 |\widetilde{-A-2,u}\rangle)=\langle u,u\rangle,
\end{equation}
where $\langle u,u\rangle$ is a bilinear and positive (over $\mathbb{R}$) pairing on the subspace of spinors satisfying $\psi_0^- u=0$. For details see Appendix B.

\subsection{Pairing on cohomology}
Now observe that (\ref{q_dual}) means that we have an induced pairing on cohomology. Recall that at picture number $A$ we have cohomologies only at ghost numbers $A+{1\over 2}$ and $A+{3\over 2}$ and that they can be realized as the spaces of representatives $\ker L_0\cap \Omega\mathcal{H}^{\perp}_{A,\downarrow}$ and $\ker L_0\cap \Omega\mathcal{H}^{\perp}_{A,\uparrow}$ respectively. So to describe the pairing on cohomology it is enough to describe the following pairing:
$$
\Big(\Omega(\mathcal{H}^{\perp}_{A,\downarrow}\oplus\mathcal{H}^{\perp}_{A,\uparrow})\cap\ker L_0\Big)\times \Big(\Omega(\mathcal{H}^{\perp}_{-A-2,\downarrow}\oplus\mathcal{H}^{\perp}_{-A-2,\uparrow})\cap\ker L_0\Big) \to \mathbb{C}.
$$
Notice that by (\ref{conj_prop}), (\ref{nsfix}) and (\ref{rfix}), the pairing (\ref{pairing_H}) has the following property: the two states $O_1|\widetilde{A}\rangle$ and $O_2 c_0|\widetilde{-A-2}\rangle$ in the NS sector or $O_1|\widetilde{A,u}\rangle$ and $O_2 c_0|\widetilde{-A-2,u}\rangle$ in the R sector can pair non-trivially only if $N^{lc}$ charges of the operators $O_1$ and $O_2$ add up to zero. Recalling the definition of $\Omega$ from Lemma 3 and using that $U$ has $N^{lc}<0$, we obtain for $\phi\in\mathcal{H}^{\perp}_{A,\downarrow}\oplus\mathcal{H}^{\perp}_{A,\uparrow}$ and $\phi'\in\mathcal{H}^{\perp}_{-A-2,\downarrow}\oplus\mathcal{H}^{\perp}_{-A-2,\uparrow}$:
\begin{equation}{
\label{to_lightcone}
(\Omega\phi,\Omega\phi')=(\phi,\phi').
}
\end{equation}
Now notice that the pairing (\ref{pairing_H}) actually pairs $\mathcal{H}^{\perp}_{A,\downarrow}$ non-degenerately to $\mathcal{H}^{\perp}_{-A-2,\uparrow}$ and $\mathcal{H}^{\perp}_{A,\uparrow}$ -- to $\mathcal{H}^{\perp}_{-A-2,\downarrow}$ (this reveals itself as an insertion of $c_0$ in (\ref{nsfix}) and (\ref{rfix})). Thus (\ref{to_lightcone}) actually implies that the cohomology at the ghost number $A+{1\over 2}$ is paired non-degenerately with the cohomology at the ghost number $-A-{1\over 2}$, and the cohomology at the ghost number $A+{3\over 2}$ -- with the cohomology at the ghost number $-A-{3\over 2}$:
\begin{align}
&\mbox{picture }-A-2\quad\quad\quad\quad\quad\quad\quad\quad\quad\mbox{picture }A\cr
&H\underbracket{^{-A-3/2}\oplus H\overbracket{^{-A-1/2}\quad\quad\quad\quad\quad H}{}^{A+1/2}\oplus H}{}^{A+3/2}.
\end{align}

\section{From bilinear pairing to Hermitian inner product on cohomology}
First we should note that it is enough to study the pairing between $H^{A+1/2}$ (at the momentum $p$) and $H^{-A-1/2}$ (at the momentum $-p$). The pairing between $H^{-A-3/2}$ and $H^{A+3/2}$ is then included automatically by replacing $A\to -A-2$.

Recall that $\mathcal{H}^{\perp}_{A,\downarrow}$ is built over the state $|\widetilde{A}\rangle$ (or over the space of states $|\widetilde{A,u}\rangle$ with $\psi_0^- u=0$ in the R sector) by exciting the oscillators $\alpha^i_{-n}, \psi^i_{-r}$, $i=2\dots9$ with $n>0$, $r> 0$, while $\mathcal{H}^{\perp}_{-A-2,\uparrow}$ is built over $c_0|\widetilde{-A-2}\rangle$ (or again $c_0|\widetilde{-A-2,u}\rangle$ with $\psi_0^- u=0$ in the R sector) by exciting the same oscillators. Thus $\mathcal{H}^{\perp}_{A,\downarrow}$ and $\mathcal{H}^{\perp}_{-A-2,\uparrow}$ are isomorphic to the light-cone spaces of (off-shell) states. Note that the subspace of spinors $u\in\textbf{32}$ satisfying $\psi_0^- u=0$ is isomorphic to the space of $Spin(8)$ spinors, the spinors in the light-cone gauge.  Since we had for the cohomology $H^{A+1/2}\cong \ker L_0\cap \Omega\mathcal{H}^{\perp}_{A,\downarrow}$ and $H^{-A-1/2}\cong\ker L_0\cap \Omega\mathcal{H}^{\perp}_{-A-2,\uparrow}$, these gave us isomorphisms of cohomologies with the light-cone spectra of the superstring at the corresponding momenta. Denote the light-cone space of \emph{on-shell} states at the momentum $p$ by $\mathcal{F}_{lc}(p)$. Denote these isomorphisms by:
\begin{align}
&\Omega_1: H^{A+1/2} \to \mathcal{F}_{lc}(p)\cr
&\Omega_2:H^{-A-1/2} \to \mathcal{F}_{lc}(-p).
\end{align}
Now we want to use these isomorphisms to define a positive Hermitian inner product on cohomology.

To define such an inner product we need to define some Lorentz-invariant \emph{antilinear} isomorphism between $H^{A+1/2}$ at momentum $p$ and $H^{-A-1/2}$ at momentum $-p$, or equivalently between $\mathcal{F}_{lc}(p)$ and $\mathcal{F}_{lc}(-p)$. This antilinear isomorphism will map between the \emph{incoming} and the \emph{outgoing} string states. Indeed, it is clear from the connection of our bilinear pairing with the two-point function on the sphere, that if we want to have a chance to define a positive inner product, we definitely have to make one string state outgoing and another one incoming. In such case conservation laws will allow pairing of two equal states.

The lowest state in the space $\mathcal{F}_{lc}(p)$ for the NS-sector is $|p\rangle$ -- the state of the space-time momentum $p$ with no oscillator excitations. In the R-sector this state is tensored with the 16-dimensional Dirac representation of $Spin(8)$, i.e. the vacuum is a subspace spanned by $|p\rangle\otimes|u\rangle$, where $|u\rangle$ is some real spinor of $Spin(8)$. We will use notation $|p,u\rangle$ as well. In any case $|p\rangle=e^{ipX}|0\rangle$, so complex conjugation reverses the sign of $p$. Since $p^{\mu}=(2\alpha')^{-1/2}\alpha_0^{\mu}$, it is natural to extend the complex conjugation operation to all the oscillators $\alpha_n^{\mu}$ and require that they change their signs under it. We define a map:
\begin{equation}
\label{antilinear}
*:\mathcal{F}_{lc}(p)\to\mathcal{F}_{lc}(-p),
 \end{equation}
 which is $\mathbb{C}$-antilinear and satisfies:
\begin{align}
\label{oscillators_a}
*\alpha_n^{i}&=-\alpha_n^{i}*\\
\label{oscillators_psi}
NS:\quad *\left(\psi^{i_n}_{-r_n}\dots\psi^{i_1}_{-r_1}|p\rangle\right)&=i^n \psi^{i_1}_{-r_1}\dots \psi^{i_n}_{-r_n}|-p\rangle\\
R:\quad *\left(\psi^{i_n}_{-r_n}\dots\psi^{i_1}_{-r_1}|p,u\rangle\right)&=\left((-1)^{\sigma+\chi(u)}i\right)^n \psi^{i_1}_{-r_1}\dots \psi^{i_n}_{-r_n}|-p,u\rangle.
\end{align}

Here we have introduced some seemingly unnatural phase factors in order to make the inner product we are about to define positive.
So for $\psi,\phi\in H^{A+1/2}$ let us define:
\begin{equation}
\label{inn_prod_coh}
\langle\phi|\psi\rangle = \Big(\psi, (\Omega_2^{-1}\circ *\circ \Omega_1)(\phi)\Big),
\end{equation}
where $(\ ,\ )$ denotes the bilinear pairing of cohomologies described in the previous section.

\textbf{Statement:} (\ref{inn_prod_coh}) is a positive Hermitian inner product on cohomology.

Let us check that this inner product is indeed positive (hermiticity follows from the antilinearity of $*$).
For the NS sector take $\psi = \Omega\alpha_{-n_1}^{i_1}\dots\alpha_{-n_k}^{i_k}\psi_{-m_1}^{j_1}\dots\psi_{-m_s}^{j_s}|\widetilde{A}\rangle$. Then 
\begin{align}
\Omega_1(\psi)&=\alpha_{-n_1}^{i_1}\dots\alpha_{-n_k}^{i_k}\psi_{-m_1}^{j_1}\dots\psi_{-m_s}^{j_s}|p\rangle\cr
* \Omega_1(\psi)&=i^{s}(-1)^k\alpha_{-n_1}^{i_1}\dots\alpha_{-n_k}^{i_k}\psi_{-m_s}^{j_s}\dots\psi_{-m_1}^{j_1}|-p\rangle\cr
&=i^s (-1)^{k+{1\over 2}s(s-1)}\alpha_{-n_1}^{i_1}\dots\alpha_{-n_k}^{i_k}\psi_{-m_1}^{j_1}\dots\psi_{-m_s}^{j_s}|-p\rangle\cr
(\Omega_2^{-1}* \Omega_1)(\psi)&=i^{s}(-1)^{k+{1\over 2}s(s-1)}\Omega\alpha_{-n_1}^{i_1}\dots\alpha_{-n_k}^{i_k}\psi_{-m_1}^{j_1}\dots\psi_{-m_s}^{j_s}c_0|\widetilde{-A-2}\rangle.
\end{align}
So we obtain:
\begin{align}
\langle\psi|\psi\rangle = i^{s}(-1)^{k+{1\over 2}s(s-1)} \Big(\alpha_{-n_1}^{i_1}\dots\alpha_{-n_k}^{i_k}\psi_{-m_1}^{j_1}\dots\psi_{-m_s}^{j_s}|\widetilde{A}\rangle,\ \alpha_{-n_1}^{i_1}\dots\alpha_{-n_k}^{i_k}\psi_{-m_1}^{j_1}\dots\psi_{-m_s}^{j_s}c_0|\widetilde{-A-2}\rangle\Big).
\end{align}

Next we use duality relations (\ref{conj_prop}) for our bilinear pairing and find that the phase factor we introduced above by hands exactly offsets the phase factors arising from using (\ref{conj_prop}). Commuting creation and annihilation operators and then using (\ref{nsfix}), we obtain $\langle\psi|\psi\rangle=n_1n_2\dots n_k>0$ (or just $1$ if there are no $\alpha_n^i$ oscillators).

The computation for the R sector is analogous. In that case we take a state $\psi = \Omega\alpha_{-n_1}^{i_1}\dots\alpha_{-n_k}^{i_k} \psi_{-m_1}^{j_1}\dots\psi_{-m_s}^{j_s}|\widetilde{A,u}\rangle$ and obtain $\langle\psi|\psi\rangle=n_1 n_2\dots n_k\langle u,u\rangle$, which is positive because $\langle \_,\_\rangle$ was a positive-defined inner product on real spinors, and adding complex conjugation turned it into positive Hermitian inner product on complexified spinors.

So after all we see that indeed we have a positive Hermitian inner product on cohomology in both sectors.

\section{Discussion}
The BRST cohomology at non-zero momentum consists of two copies of the light-cone spectrum at two adjacent ghost numbers. This was proven above by straightforward application of the spectral sequence for the light-cone filtration. However, it did not give an explicit realization of the physical states.

The method from Section 6 provided such a realization, describing the BRST cohomology as isomorphic to a certain subspace in the space of BRST-closed states (this is essentially the same approach as in the Hodge theory). One can show (as in \cite{Pol} for the bosonic case) that the part of the space $\ker(S+U)\cap\ker L_0$ annihilated by $b_0$, i.e. $\tilde{\mathcal{H}}^{\perp}_{A+1/2}\cap\ker L_0$, is actually isomorphic to the OCQ space of states (if the picture number $A$ is canonical, i.e. $-1$ for the NS sector or $-1/2$ for the Ramond sector). 

The key idea is to consider grading by $N'=2N^- + N_b + N_c + N_{\gamma} + N_{\beta}$, where $N^-$ is the total number of $-$-excitations above the vacuum $| -1\rangle$ or $|-{1\over 2},u\rangle$ (we are working in the canonical pictures now), and $N_b, N_c, N_{\gamma}, N_{\beta}$ are the numbers of $b, c, \gamma, \beta$ excitations. One can easily check that the operator $R$ from (\ref{R_operator}) has $N'=-1$. Also one can notice that $Q_0 + Q_{-1}$ has at most terms with $N'=1$, and so $U=\{R,Q_0+Q_{-1}\}$ doesn't increase the $N'$ number.  Thus, the map $\Omega$ doesn't increase the $N'$-number either. 

But from the structure of $\ker S$, we know that $\ker S$ has $N'=0$ (for the canonical pictures again). Thus, since $N'\ge 0$, we deduce that the space $\tilde{\mathcal{H}}^{\perp}_{A+1/2}\cap\ker L_0$ actually has no ghost and $''-''$ excitations. Such a state has the form $|matter\ state\rangle\otimes|ghost\ vacuum\rangle$ and thus is the OCQ-type state. To finish the proof, one checks that every OCQ state $|matter\ state\rangle$ gives rise to the BRST class represented by  $|matter\ state\rangle\otimes|ghost\ vacuum\rangle$ and then checks that this map sends OCQ-null states to BRST-exact states and that pre-image of every BRST-exact state is OCQ-null. The existence of the positive inner product on cohomology is actually used in this proof (see \cite{Pol} for the details on the bosonic case).

We should also note that the question of the zero-momentum cohomology is beyond the scope of this paper. It was proven in \cite{Horowitz:1988ip} using the picture-changing operators that the absolute cohomologies are isomorphic at different picture numbers. Their proof includes zero-momentum case as well. However the question of computing relative cohomology at zero momentum for arbitrary picture number remains open.
\section*{Acknowledgments}
The author is grateful to E.Witten for valuable discussions and for suggesting the problem. The author also would like to thank A.Losev, V.Mikhaylov and I.Klebanov for useful discussions.

\section*{Appendix}
\appendix
\section{$\mathbb{Z}_2$-grading}
In the NS sector the state $|A\rangle=|p\rangle\otimes|\downarrow\rangle\otimes|A\rangle_{\beta\gamma}$ has $\mathbb{Z}_2$-degree $(-1)^{A+1}$. The reason is that the state $|\downarrow\rangle$ is odd (it corresponds to the odd field $c(z)$), the state $|p\rangle$ is even (it corresponds to $e^{ipX}$), and the state $|A\rangle_{\beta\gamma}$ corresponds to the operator $\Theta^A(z)$, which conventionally is postulated to have the degree $(-1)^A$ (this operator is usually bosonized as $e^{A\phi}$). 

In the R sector, assigning some $\mathbb{Z}_2$-degree to the state $|A,u\rangle=|p\rangle\otimes|u\rangle\otimes|\downarrow\rangle\otimes|A\rangle_{\beta\gamma}$ is a more subtle question. The vector $|p\rangle$ is still even and $|\downarrow\rangle$ is still odd. However $|u\rangle$ and $|A\rangle_{\beta\gamma}$ correspond to $u_L^{\alpha}\Sigma_{\alpha}(z)+u_{R,\alpha}\Sigma^{\alpha}(z)$ and $\Theta^A(z)$ respectively, where $\Sigma_{\alpha}(z), \Sigma^{\alpha}(z)$ are the spin-fields of opposite chirality for $\psi^{\mu}$'s, and $\Theta^A(z)$ is a spin-field for $\beta\gamma$-ghosts. Neither $\Sigma_{\alpha}(z), \Sigma^{\alpha}(z)$ nor $\Theta^A(z)$ are bosonic or fermionic. 

Indeed, from the OPE of spin-fields (\cite{Friedan:1985ge}):
\begin{align}
\label{spin_fields}
\Sigma^{\alpha}(z)\Sigma_{\beta}(w)&\sim (z-w)^{-5/4}\left(\delta_{\beta}^{\alpha} + O(z-w)\right)\cr
\Sigma_{\alpha}(z)\Sigma_{\beta}(w) &\sim (z-w)^{-3/4}\left({1\over\sqrt{2}}\Gamma^{\mu}_{\alpha\beta}\psi_{\mu}(z)+O(z-w)\right)\cr
\psi^{\mu}(z)\Sigma_{\alpha}(w) &\sim {1\over (z-w)^{1/2}}\left({1\over\sqrt{2}}\Gamma^{\mu}_{\alpha\beta}\Sigma^{\beta}(w)+O(z-w)\right),
\end{align}
we see that the product of two spin fields of the same chirality is fermionic, the product of two spin fields of the opposite chirality is bosonic, and the product of the spin field of definite chirality with the field $\psi^{\mu}$ has the opposite chirality. It is quite clear that it is impossible to consistently introduce $\mathbb{Z}_2$-grading into this algebra such that $\psi^{\mu}$ is still fermionic. However $\mathbb{Z}_4$-grading works well. One can still say that $\psi^{\mu}$ has degree $-1$, while $\Sigma_{\alpha}(z)$ has degree $-(-1)^{\sigma} i$, and $\Sigma^{\alpha}$ has degree $(-1)^{\sigma} i$. Here $\sigma=0\mbox{ or }1$, -- there is no way to canonically fix the sign of $(-1)^{\sigma}$. 

As for the field $\Theta^A(z)$, it also cannot have just a $\mathbb{Z}_2$-degree. One reason for that is that its bosonization is as usual $e^{A\phi}$ with half-integer $A$. In the NS sector, bosonization was the same but with integer $A$, and the degree there was chosen to be $(-1)^A$, which means that it is inconsistent to set the degree to just $+1$ or $-1$ for the R sector. Another, more important reason is that $\Theta^A(z)$ should offset the unusual statistics of $\Sigma_{\alpha}(z)$ and $\Sigma^{\alpha}(z)$. The consistent way to define grading here is to say that $\Theta^A(z)$ has degree $e^{i\pi A}=-i (-1)^{A+1/2}$. We do not have to introduce any arbitrary unfixed sign here, since we already have introduced $\sigma$ above.

So now the product $\Sigma_{\alpha}(z)\Theta^A(z)$ has the degree $(-1)^{A+\sigma+3/2}$, and $\Sigma^{\alpha}(z)\Theta^A(z)$ has the degree $(-1)^{A+\sigma+1/2}$. These are $\pm 1$, which means that now we can introduce $\mathbb{Z}_2$-grading into the space $\mathcal{H}^A$ consistently even in the R sector.

\textbf{Remark:} Note that in the Type IIB and Type IIA theories, different choices of $\sigma$ for the right-moving R-states are required. Indeed, in the IIB we have a sector $(NS+,R+)$, which describes space-time fermions. Thus the R+ sector vertex operators (where sign denotes the GSO-degree, which in the canonical picture coincides with the chirality of the Ramond vacuum) must be the world-sheet fermions here, which fixes $\sigma=0$ for the right-moving R-sector states in the Type IIB theory. In the Type IIA we have, on the contrary, the $(NS+,R-)$ sector, which still should describe fermions, which means that in that case we should fix $\sigma=1$ for the right-moving R-sector states. In both theories we fix $\sigma=0$ for the left-moving states. This guarantees the correct spin-statistics connection for the fields that survive the GSO projection.

Now we can find the consistent $\mathbb{Z}_2$ degree for the state $|A,u\rangle=|p\rangle\otimes|u\rangle\otimes|\downarrow\rangle\otimes|A\rangle_{\beta\gamma}$. The chirality of the spin field is fixed to be $+1$ for $\Sigma_{\alpha}(z)$ and $-1$ for $\Sigma^{\alpha}(z)$. Let us denote the chirality of the spinor $|u\rangle$ by $(-1)^{\chi(u)}$. Then, counting the degrees, the $\mathbb{Z}_2$-degree of $|A,u\rangle$ is $(-1)^{A+\sigma+1/2+\chi(u)}$.

 $\mathbb{Z}_2$-grading of any other state in $\mathcal{H}^A$ is determined by counting bosonic and fermionic excitations above $|A\rangle$ or $|A,u\rangle$. 

By counting excitations one can determine the degree of the state $|\widetilde{A}\rangle$ (see (\ref{truevac})) in the NS sector to be just $+1$. The state $|\widetilde{A,u}\rangle$ has the degree $(-1)^{\sigma+\chi(u)}$.
\section{Fixing normalization in the Ramond sector}
The Ramond sector is a little more subtle. First, let us consider separately the pairing for spinors. We are interested in $(u,v)$, where $u,v\in \textbf{32}$. From (\ref{spin_fields}), we know the two-point functions of the spin fields:
\begin{align}
\langle\Sigma_{\alpha}(z)\Sigma_{\beta}(0)\rangle&=\langle\Sigma^{\alpha}(z)\Sigma^{\beta}(0)\rangle=0\cr
\langle\Sigma^{\alpha}(z)\Sigma_{\beta}(0)\rangle&=\langle\Sigma_{\beta}(z)\Sigma^{\alpha}(0)\rangle={\delta^{\alpha}_{\beta}\over z^{5/4}}.
\end{align}
This shows how to pair spinors: 
\begin{equation}
(u,v)=u_L^{\alpha}v_{R,\alpha} + u_{R,\alpha}v_L^{\alpha}.
\end{equation}
Now recall that the spinor $u$ appearing in $|\widetilde{A,u}\rangle$ is required to satisfy $\psi_0^- u=0$ by definition (\ref{truevac_R}). If we take two such spinors $u$ and $v$ satisfying $\psi_0^-u=\psi_0^-v=0$, then one can write $v=\psi_0^- w$ and use the conjugation property:
$$
(u,v)=(u,\psi^-_0 w)\ \propto\ (\psi_0^-u,w)=0.
$$
However, we can define another pairing:
\begin{equation}
\label{newpairing}
\langle u,v\rangle= -(u,\Gamma\psi_0^+v),
\end{equation}
where $\Gamma$ is a chirality matrix. This pairing will be non-trivial on the subspace of spinors satisfying the $\psi_0^- u=0$ condition. Since $\textbf{32}$ is a real representation, $\langle u,v\rangle$ gives some inner product on the real subspace. This inner product is actually positive definite. The reason is that $-\Gamma\psi_0^+$ acts as a charge conjugation matrix (in the Majorana basis) on $v$. On the other hand, the matrix of the bilinear pairing is the same charge conjugation matrix. So if we fix the Majorana basis, in the matrix notations our new pairing on spinors is just:
$$
\langle u,v\rangle = u^T C C v = u^T v,
$$
where we used that $C^2=1$ for $C=-\Gamma \Gamma^0$, which is a charge conjugation matrix in the Majorana basis. The last expression clearly gives an inner product which is positive over the reals.

Now when we pass to the full Fock space, we obtain $(|A,u\rangle, c_0\psi_0^+|-A-2,v\rangle)=\pm\langle u,v\rangle$. To fix the sign, we require that for every $u$ which is annihilated by $\psi_0^+$, we have:
\begin{equation}
R:\quad (|\widetilde{A,u}\rangle, c_0 |\widetilde{-A-2,u}\rangle)=\langle u,u\rangle.
\end{equation}
Note that in the R sector one (and only one) of the states $|\widetilde{A}\rangle$ and $|\widetilde{-A-2}\rangle$ has the $\psi_0^+$ excitation, so the above condition makes sense.

\end{document}